\documentclass[conference]{IEEEtran}
\IEEEoverridecommandlockouts
\pdfoutput=1

\usepackage{tikz}
\usepackage{textcomp}
\usepackage{hyperref}
\usepackage{lipsum}

\newcommand\copyrighttext{%
  \footnotesize \textcopyright 2021IEEE. Personal use of this material is permitted.
  Permission from IEEE must be obtained for all other uses, in any current or future
  media, including reprinting/republishing this material for advertising or promotional
  purposes, creating new collective works, for resale or redistribution to servers or
  lists, or reuse of any copyrighted component of this work in other works.
  }
\newcommand\copyrightnotice{%
\begin{tikzpicture}[remember picture,overlay]
\node[anchor=south,yshift=10pt] at (current page.south) {\fbox{\parbox{\dimexpr\textwidth-\fboxsep-\fboxrule\relax}{\copyrighttext}}};
\end{tikzpicture}%
}

\usepackage{cite}
\usepackage{amsmath,amssymb,amsfonts}
\usepackage{algorithmic}
\usepackage{graphicx}
\usepackage{textcomp}
\usepackage{xcolor}
\usepackage{caption}
\usepackage{subcaption}
\usepackage{algorithm} 
\usepackage{algorithmic}
\usepackage{multirow}
\usepackage{booktabs}

\usepackage{cleveref}
\crefname{section}{§}{§§}
\Crefname{section}{§}{§§}

\usepackage{xcolor}  
\usepackage{soul}

\def\BibTeX{{\rm B\kern-.05em{\sc i\kern-.025em b}\kern-.08em
    T\kern-.1667em\lower.7ex\hbox{E}\kern-.125emX}}
\begin{document}

\title{Learning-to-Dispatch: Reinforcement Learning Based Flight Planning under Emergency 
\thanks{
* This work was supported by the Center for Advanced Transportation Mobility (CATM), USDOT Grant \#69A3551747125.
}
\thanks{
\textsuperscript{1}Kai Zhang, Yupeng Yang, Chengtao Xu and Houbing Song are with the Security and Optimization for Networked Globe Laboratory (SONG Lab), Embry-Riddle Aeronautical University, Daytona Beach, FL 32114 USA. 
{\fontfamily{qcr}\selectfont
\{zhangk3, yangy5, xuc3\}@my.erau.edu, h.song@ieee.org}
}
\thanks{
\textsuperscript{2}Dahai Liu is with the College of Aviation, Embry-Riddle Aeronautical University, Daytona Beach, FL 32114 USA.
{\fontfamily{qcr}\selectfont
dahai.liu@erau.edu
}}
}

\author{\IEEEauthorblockN{Kai Zhang\textsuperscript{1} , Yupeng Yang\textsuperscript{1}, Chengtao Xu\textsuperscript{1}, Dahai Liu\textsuperscript{2} and Houbing Song\textsuperscript{1}}
}

\maketitle
\copyrightnotice

\begin{abstract}
The effectiveness of resource allocation under emergencies especially hurricane disasters is crucial. However, most researchers focus on emergency resource allocation in a ground transportation system. 
In this paper, we propose Learning-to-Dispatch (L2D), a reinforcement learning (RL) based air route dispatching system, that aims to add additional flights for hurricane evacuation while minimizing the airspace's complexity and air traffic controller's workload. Given a bipartite graph with weights that are learned from the historical flight data using RL in consideration of short- and long-term gains, we formulate the flight dispatch as an online maximum weight matching problem. Different from the conventional order dispatch problem, there is no actual or estimated index that can evaluate how the additional evacuation flights influence the air traffic complexity. Then we propose a multivariate reward function in the learning phase and compare it with other univariate reward designs to show its superior performance.
The experiments using the real-world dataset for Hurricane Irma demonstrate the efficacy and efficiency of our proposed schema.
\end{abstract}


\section{Introduction}

Natural disasters such as floods, earthquakes, and wildfires occur around the world almost every year and cause loss of life or damage property \cite{smith2013us}. Hurricane is the most common flooding disaster in the Southeastern United States and it has killed or injured many populations. Therefore, how to design an effective evacuation plan before a hurricane to reduce causalities becomes a significant problem for emergency planning and management \cite{zhu2019modeling, tri2020studying}. 

\begin{figure}
    \centering
    \includegraphics[scale = .34]{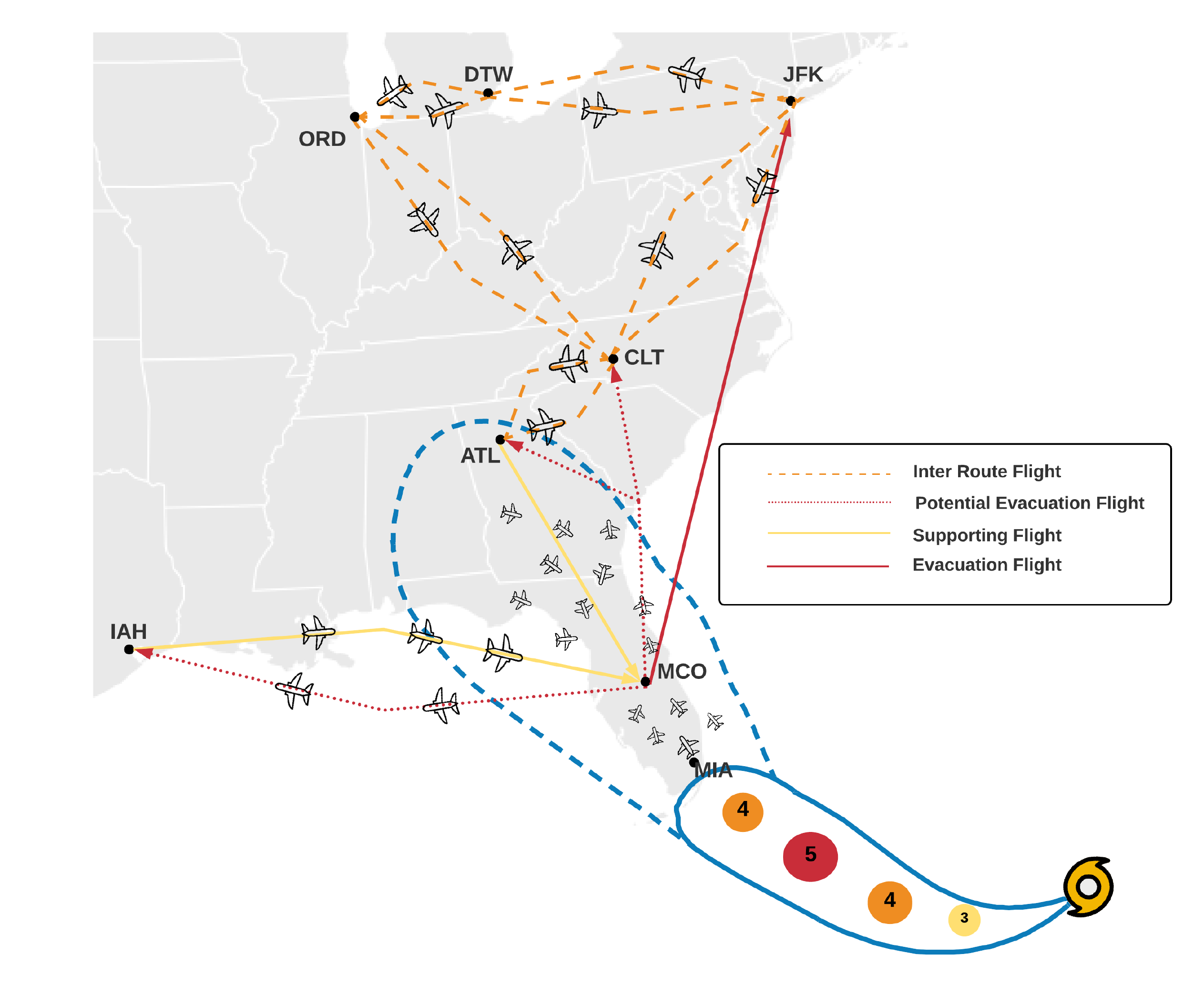}
    \caption{Flight Dispatch under Emergency.}
    \label{fig:evacuation}
\end{figure}

According to the Hurricane Irma Local Report, approximately 6+ million South Floridians joined in evacuating. Although most of them began paying very close attention to Irma’s forecast up to a week and left early, they still encountered significant traffic jams upstate.
To deal with this issue, recent studies \cite{yi2017optimization, zhu2020estimating, dulebenets2019exact} focus on vehicle routing or route choice problem during hurricane evacuation considering both dynamic evacuation demand \cite{yin2014agent} and hurricane characteristics. In summary, they estimated evacuee behavior and assessed road accessibility using statistical models, then applied dynamic traffic assignment approaches \cite{lv2015managing} to solve for the paths and travel times of each driver. Indeed, the practical application of these researches is in doubt because people do not usually follow the recommended route, especially in emergencies. The indeterminate driving behaviors bring unknowable disturbance to the system. Consequently, taking a flight to evacuate before the hurricane is a better option since air traffic control (ATC) is a centralized system in which "drivers" and paths are managed by the air transportation authorities. However, to the best of our knowledge, there has been no literature thus far that tries to address the flight dispatch during a hurricane evacuation. On the other hand, recent advances in Internet of Things and computing power made large-scale traffic data be collected and analyzed in the modern intelligent transportation system (ITS) \cite{song2017smart, lv2014traffic, sun2016internet, 7926913,dartmann2019big}. As the transportation management and control becomes more data driven \cite{zhu2018big, jiang2021spatial, zhang2020spatio}, machine learning has drawn a lot of academic interest in the application of disaster evacuation planning \cite{zhao2020modelling, burris2015machine, song2013modeling, reynard2019harnessing}. However, previous work purely focused on classification tasks, such as driver's route choice \cite{zhao2020modelling, song2013modeling}, contraflow activation \cite{burris2015machine} and human's sentiments \cite{reynard2019harnessing}, while they naturally ignore the potential application of data-driven evacuation planning via machine learning. 

In the paper, we propose Learning-to-Dispatch (L2D), a reinforcement learning (RL) based flight dispatch system that aims to maximize the air traffic capacity, and meanwhile minimize the potential negative impact of additional flights on the air route network. To accomplish this, we first generate synthetic flight transactions and model the dynamics of the air route network during hurricane evacuation based on the proposed reward function which balances the trade-off between evacuation efficiency and airspace capacity. The flight planning is then formulated as a matching problem, where each individual decision of matching an aircraft to an evacuation flight is based on the instant reward for the aircraft serving this flight and the impact of this decision in the future. 


The contribution of this work is summarized as follows:
\begin{itemize}
    \item To our knowledge, we firstly propose an effective flight dispatch system for supporting disaster evacuation. The system considers both instance requirements of evacuation efficiency and the expected future air traffic complexity.
    \item We present an elaborated reward function. The experimental results prove that it results in a more balanced dispatch plan in terms of flight length, flight elapsed time, and air traffic controller's workload, which is likely to play a vital role in the future optimization of evacuation flight dispatching research.
\end{itemize}

The remainder of this paper is structured as follows. Section \ref{sec:problem} formulates the flight dispatch problem. The L2D system design is detailed in Section \ref{sec:system}, including synthetic data generation and model construction. Experiments are then described in Section \ref{sec:evacuation}, followed by the conclusion of the paper in Section \ref{sec:conclusion}.


\section{Problem Formulation} \label{sec:problem}

Figure. \ref{fig:evacuation} shows the basic idea behind the L2D system. More specifically, flights are categorized by task type: (a) inter route is the scheduled flight which serves regular trips; (b) evacuation route is the extra flight or charter flight for hurricane evacuation; (c) supporting route can be considered a kind of evacuation route but it is not "direct" one whose departure is in the evacuation zones. Given the estimated arrival time of hurricane $T_h$, the flight dispatching aims to complete the urgent egress of people away from a city at least $T$ hours before a hurricane approaches. To achieve this goal via a coordinated and optimized way, we should take two principle problems into account: (a) \textit{How to measure the long-term increase of air traffic complexity caused by the evacuation flight?}, and (b) \textit{How to allocate available resources on basis of future complexity estimation?}

\begin{figure}
    \centering
    \includegraphics[scale = .45]{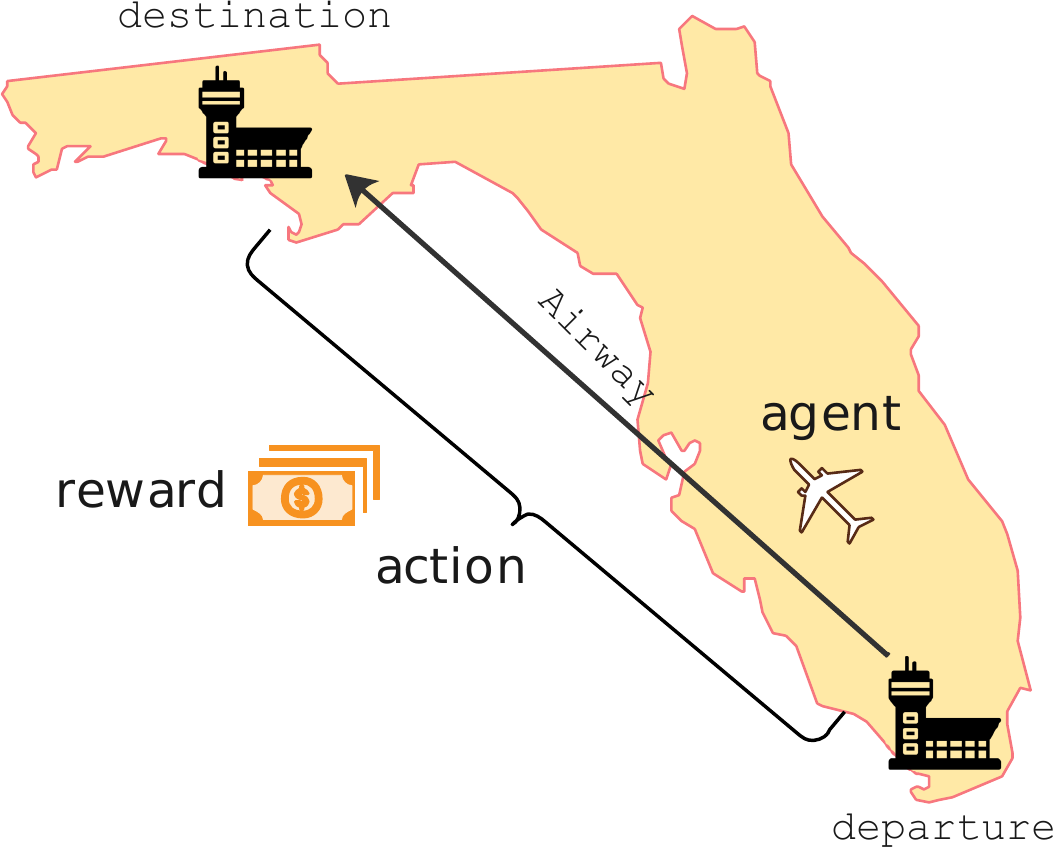}
    \caption{MDP formulation in the L2D. An agent represents an aircraft/pilot.The action in flight dispatch represents assigning the aircraft/pilot to serve a particular evacuation flight.}
    \label{fig:mdp}
\end{figure}

To solve these issues, we model flight dispatch as a sequential decision problem, say fully observable Markov decision process (MDP), in which there is an embedded decision node of matching aircraft-airway at each stage. The key terms of MDP we build in the paper are discussed in the following paragraphs and illustrated in Figure \ref{fig:mdp}.

\subsection{MDP Definition}

An MDP is defined by a tuple $\mathcal{M}=\langle \mathcal{S}, \mathcal{A}, \mathcal{P}, \mathcal{R}, \gamma \rangle$, where $\mathcal{S}, \mathcal{A}, \mathcal{P}, \mathcal{R}, \gamma$ are the set of states, set of actions, transition probability function, reward function and a discount rate, respectively. The purpose of the agent is to maximize the long-run reward $G_{t} = \sum_{k=0}^T \gamma^k R_{t+k+1}$ it receives. There are two strategies to model an agent: (a) To model the dispatch system as an agent; and (b) To model the individual aircraft preparing for serving evacuation as an agent. For the first strategy, it is hard to define components in an MDP in an appropriate manner. As an example, we have to cope with the large combinatorial action spaces. Therefore, we adopt the second strategy. However, it still exists a major problem that the state spaces and action spaces are changing for each agent at different stages. Here we use a generalized setting. Specifically, we regard each agent as the same so that each agent performs the same policy $\pi$. Other components in an MDP can be defined as follows:


\begin{itemize}
    \item \textbf{State} $\mathcal{S}$ is determined by a spatial-temporal vector $\langle g, t \rangle$ where $g$ is the airport index where the aircraft is located and $t$ is the time index. 
    \item \textbf{Action} $\mathcal{A}$ is defined as a set $\{0, 1\}$ where $0$ and $1$ represent that the aircraft is assigned to a flight or not in a time slot, respectively.
    \item \textbf{State Transition} $\mathcal{P}$ depends on the estimated future information such as flight delay and air flow. We use a naive way to estimate these information for simplicity that we assume the relevant variables are distributed normally with the calculated means and variances based on historical records.  
    \item \textbf{Reward} $\mathcal{R}$ is defined as an index with hybrid factors which we think can reflect the influence of an evacuation flight to the corresponding airports. Details are seen in \cref{sec:value_function}, the definition of ATC impact factor (ATCI).
    \item \textbf{Discount rate} $\gamma$ determines how important future rewards to the current state. We set $\gamma = 0.9$ in the paper.
\end{itemize}

\subsection{Dynamics of the Airspace Complexity} \label{sec:value_function}

We use the ATCI to measure the airspace complexity which is defined as follows:

\begin{equation}
    \text{ATCI} = \frac{100}{\lambda \times w + \mu \times et + \nu \times d}
\end{equation}
where $w$ is an indicator for air traffic controller's workload. 
$et$ denotes the elapsed time of a flight, $d$ represents the distance, while $\lambda, \mu, \nu$ are hyperparameters. The details of how to calculate indicators and select hyperparameters are described in \cref{sec:simulation}. 

Recall the MDP definition, an agent performs a policy $\pi$ in an environment (the air route network in our problem setting) and tries to maximize its gain. To get the optimal policy, one path is to learn the value function $V_{\pi}(s) = \mathbb{E}_{\pi} [G_t | S_t = s]$ that estimate how good it is for an agent does a given action in a given state. Logically, it is clear that $V_{\pi}(s)$ measures the dynamics of airspace complexity since $V_{\pi}(s) \sim \mathrm{ATCI}$ and ATCI is related to factors of airspace complexity. In other words, the learned value function captures the spatio-temporal patterns of the air route network that can further used to dispatch evacuation flights.





\subsection{Flight Dispatching under Emergency} \label{sec:matching}

Now we assume there are total $m$ available aircraft and $n$ airports at a time slot $t$ that are able to assist people to evacuate. The goal of flight dispatching is to find the best match between potential aircraft and potential evacuation flights that are defined mathematically as follows:

\begin{itemize}
    \item \textbf{Potential Aircraft} is denoted by a set of quadruple $\langle I_a, S_a, A_a, C_\rangle$, where $I_a, S_a, A_a, C_a$ are the index of aircraft, current state, availability index and expected cumulative cost respectively.
    \item \textbf{Potential Evacuation Flights} is represented by a set of quintuple $\langle I_o, S_D, A_o, C_o, S_A \rangle$, where $I_o, S_D, A_o, C_o, S_A$ are the index of flight, departure state, availability index, expected ATC impact, and expected arrival state respectively. Note that in the evacuation task, the airway could be a single itinerary from city $\mathcal{C}$ to the arrival city or multi-way itinerary. For example, an aircraft takes off in the city $\mathcal{C}_D$, and flies to $\mathcal{C}$ to support the evacuation and then goes to the other city $\mathcal{C}_A$.
\end{itemize}

The best match means that at each time step, potential airways are paired with the unoccupied aircraft resulting in the maximum expected dispatch gain or minimum expected air traffic complexity.
Formally, this task's objective function is written as follows:

\begin{equation}
\begin{aligned}
\underset{a_{i j}}{\operatorname{argmax}} & \sum_{i=0}^{m} \sum_{j=0}^{n} W(i, j) a_{i j} \\
\text { s.t. } & \sum_{i=0}^{m} a_{i j}=1, \quad j=1,2,3 \ldots, n \\
& \sum_{j=0}^{n} a_{i j}=1, \quad i=1,2,3 \ldots, m
\end{aligned}
\label{eq:dispatch}
\end{equation}
where $a_{i j} = 1$ if airway $j$ is assigned to aircraft $i$ while $a_{i j} = 0$ if airway $j$ is not assigned to aircraft $i$. $W(i,j)$ is a function that could point out the long-term return or cost if an airway is dispatched.

\begin{figure}
    \centering
    \includegraphics[scale = .6]{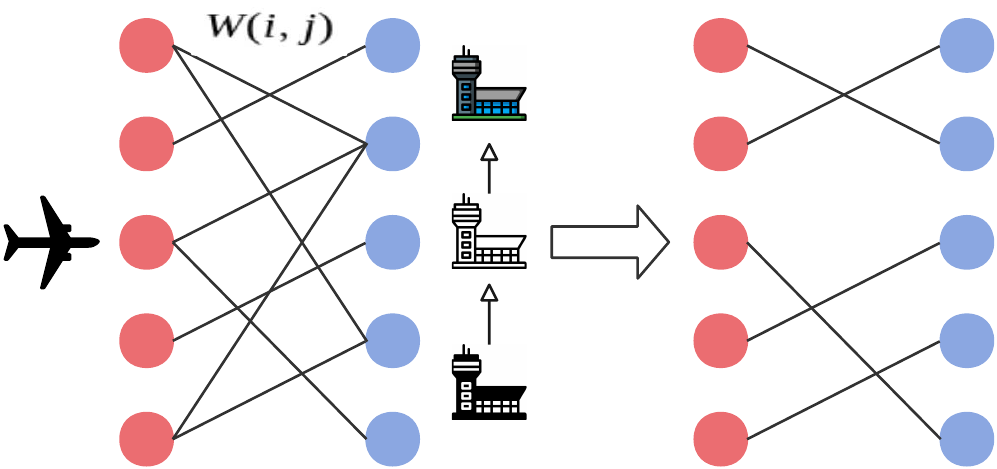}
    \caption{An aircraft-airway maximum-weight matching example, in which $i$ is the aircraft index while $j$ is the airway index. The weights of each potential connection are obtained by the estimated value functions.}
    \label{fig:match}
\end{figure}

The Eq. \ref{eq:dispatch} can be further defined as a bipartite graph matching problem with weighted edges as shown in Figure. \ref{fig:match}. The weights here are learned by RL. To put it another way, the weights come from the value function. We will explain how to obtain the value function in \cref{sec:weighted}. 

\section{System Design} \label{sec:system}
\begin{figure}
    \centering
    \includegraphics[scale = .6]{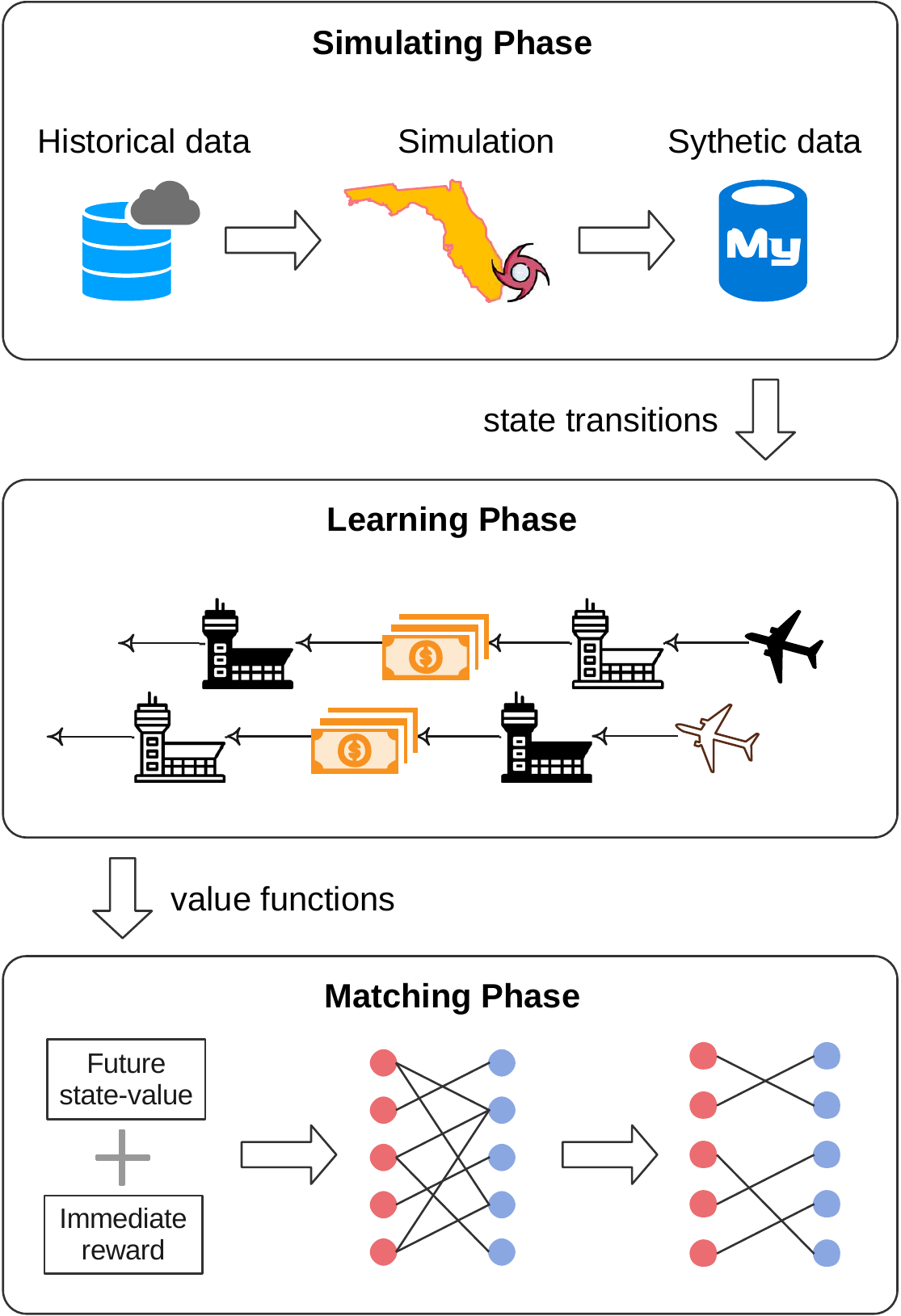}
    \caption{Overview of L2D system.}
    \label{fig:system}
\end{figure}

L2D system consists of three modules: offline simulating, offline learning, and online matching, as illustrated in Figure. \ref{fig:system}. Although it is computationally expensive for learning the value function, it could be completed before disasters approach. Thus, offline process is not time-sensitive in practice.

\subsection{Generating Simulated Evacuation Flight Transactions} \label{sec:simulation}

The flight transactions in the historical database record only a series of one-way itineraries such as from Miami to New York City. However, in the context of disaster evacuation, an evacuation airway could be a multi-way itinerary. Thus we generate simulated evacuation flight transactions based on the statistical properties from the historical real-world data, which will be utilized to learn the value functions later. The progress of producing synthetic transactions is described in Algorithm \ref{alg:generator}. In each episode, we randomly select departure and arrival states including airport and time from the historical flight dataset, and treat the availability index as either 0 (not available) or 1 (available). 

At a time $t$, workload $w^k = air^k/atc^k$, time $et^k$, and distance $d^k$ could measure a one-way flight as well as a multi-way flight. Let's see an example to make the process clearer. There is a transaction from $k_1$, supporting $k_2$ to transport people to $k_3$, therefore the $w^k_t$ should be the accumulated workload when an aircraft takes off or lands in these three airports at the corresponding time step, which is calculated as $w^k_t = w^{k_1}_{t_1} + w^{k_2}_{t_2} + w^{k_3}_{t_3}$ where $t_1$ represent the time of departure and $t_2, t_3$ represent the estimated time of arrival.

Algorithm \ref{alg:hyper_selection} is the hyperparameter selection strategy for computing ACTI. The intuition behind the hyperparameter selection algorithm is to let every variable be negatively correlated with the ATCI as much as possible so that the reward function could not be heavily biased toward either factor.

\subsection{Building Weighted Aircraft-Airway Graph} \label{sec:weighted}

We adopt the advantage function trick in \cite{xu2018large} to reduce the computational complexity by removing connections between aircraft and the "idle" action. The advantage function measures what is the advantage of selecting a certain action in a certain state compared to expected reward of all possible actions in that state. Mathematically, the advantage function is defined as:

\begin{equation} \label{eq:advantage}
    A_{\pi} (i,j) = Q_{\pi} (i,j) - V_{\pi} (s_i) 
\end{equation}
where $Q_{\pi} (i,j)$ is the action-value function of aircraft $i$ performing action of serving an evacuation transaction $j$ or "do nothing". The value of this function shows the expected return taking action $a$ in state $s$ under policy $\pi$. 




According to Eq. \ref{eq:dispatch}, we denote the case of $i = 0$ or $j = 0$ as the situation that the airway is not assigned to any aircraft. Let $s_i$ and $s_i'$ denote the current state of the aircraft $i$, and the final state when an evacuation flight transaction is completed, the state-value function is thus obtained with dynamic programming based on Temporal-Difference (TD) method \cite{sutton2018reinforcement} as described in Algorithm \ref{alg:dp}. Therefore, we can further rewrite Eq. \ref{eq:advantage} in the following:

\begin{equation} \label{eq:advantage_detail}
    A_{\pi} (i,j) = \gamma^{\Delta t_j} V(s_{ij}') + R_j - V(s_i)
\end{equation}

Now edge weight of each potential evacuation flight can be calculated using Eq. \ref{eq:advantage_detail} to build the weighted bipartite graph. If $Q_{\pi} (i,j) \leq V_{\pi} (s_i)$, we let the weight between $i$ and $j$ be zero to reduce the potential pairs, then to reduce the compuatational complexity. 

\subsection{Matching Maximum-Weighted Connections}


The aforementioned bipartite graph contains two sets of nodes -- aircraft and airway, and the edge between them has a weight of $A_{\pi}(i,j)$. We adopt the widely used Hungarian algorithm to finds a maximum matching \cite{bruff2005assignment}. 

\begin{algorithm}
\caption{Evacuation Flight Transaction Productor}
\begin{algorithmic}[1]
\renewcommand{\algorithmicrequire}{\textbf{Input:}}
\renewcommand{\algorithmicensure}{\textbf{Output:}}
\REQUIRE Historical flight data $\mathbf{F}$ and human-factor data $\mathbf{H}$; pre-defined number of transactions $\mathbf{N}$. The airports $k$ and time $t$ are extracted randomly from the historical datasets in each iteration.
\ENSURE Synthetic airway set $\{\langle I_o, S_D, A_o, C^k_t, S_A \rangle\}$ 
\FOR {$i = 0$ to $i = \mathbf{N}$}
    \STATE Generate a potential airway tuple  $\langle I_o, S_D, A_o, S_A \rangle _i$ but without the ATCI.
    \STATE $atc^k_t \leftarrow $ number of air traffic controllers $\times$ the corresponding training time
    \STATE $air^k_t \leftarrow $ the average of delay $\times$ the number of flights
    \STATE $et^k_t \leftarrow $ sample from the historical elapsed time distribution complying with Gaussian
    \STATE $d^k_t \leftarrow$ flight length
    \STATE $C^k_t = 100 / [\lambda \times (air^k_t / atc^k_t) + \mu \times et^k_t + \nu \times d^k_t]$
    \STATE Take the $C^k_t$ as the ATCI in the airway tuple to construct $\langle I_o, S_D, A_o, C^k_t, S_A \rangle _i$.
\ENDFOR
\end{algorithmic}
\label{alg:generator}
\end{algorithm}

\begin{algorithm}
\caption{Hyperparameter Selection}
\begin{algorithmic}[1]
\renewcommand{\algorithmicrequire}{\textbf{Input:}}
\renewcommand{\algorithmicensure}{\textbf{Output:}}
\REQUIRE A pre-defined hyperparameter candidate set \{$\mathcal{H}_i$\} where $i$ is the index of each candidate, $i \in [0, N_{hyper}]$.
\ENSURE An optimal hyperparameter tuple $\langle \lambda^*, \mu^*, \nu^* \rangle$.
\\ \text{Initialize $\rho_{hyper}$ and $\mathcal{H}^* = \langle \lambda^*, \mu^*, \nu^* \rangle$ of all zeros.}
\FOR {$i = 0$ to $i = N_{hyper}$}
    \STATE $\rho^w_i, \rho^d_i, \rho^t_i \leftarrow$ calculate the Spearman rank-order correlation coefficient of ATCI and workload, flight length, elapsed time in $\mathcal{H}_i$, respectively.
    \IF {$\rho_i \leq 0.4$ for each coefficient}
        \STATE  $\rho^{neg}_i = |\rho^w_i + \rho^d_i + \rho^t_i|$
            \IF {$\rho^{neg}_i > \rho_{hyper}$}
                \STATE $\rho_{hyper} \leftarrow \rho^{neg}_i$, $\mathcal{H}^* \leftarrow \mathcal{H}_i$ 
            \ENDIF
    \ENDIF
\ENDFOR
\end{algorithmic}
\label{alg:hyper_selection}
\end{algorithm}

\begin{algorithm}
\caption{State-value Lookup Generator}
\begin{algorithmic}[1]
\renewcommand{\algorithmicrequire}{\textbf{Input:}}
\renewcommand{\algorithmicensure}{\textbf{Output:}}
\REQUIRE Collect state transitions $\{\langle s_i, a_i, r_i, s_i' \rangle\}$ from synthetic evacuation transactions . State is represented as a spatio-temporal tuple: $s_i = \langle g_i, t_i \rangle$,  $s_i' = \langle g_i', t_i' \rangle$. $T$ is the time of last state in the finite-horizon MDP.
\ENSURE Value function $V(s)$ for all states.
\\ \text{Initialize $V(s)$ and $N(s)$ of all zeros}
\FOR {$t = T-1$ to $t = 0$}
    \STATE $\{\langle s_i, a_i, r_i, s_i' \rangle\}^t \leftarrow$ A subset of state transitions in which $t_i = t$
    \FOR {each instance in the subset}
        \STATE $N(s_i) \leftarrow N(s_i) + 1$
        \STATE TD error: $\delta_i \leftarrow \gamma^{\Delta t (a_i)} V(s_i') + R(a_i) - V(s_i)$
        \STATE $V(s_i) \leftarrow V(s_i) + \frac{1}{N(s_i)} \delta_i$
    \ENDFOR
\ENDFOR
\end{algorithmic}
\label{alg:dp}
\end{algorithm}

In the scenario of online evacuation flight dispatching, L2D collects all available aircraft and potential airways at every time step to build aircraft-airway pairs with weights that are estimated using Eq. \ref{eq:advantage_detail}, and afterward the matching is solved by Hungarian algorithm. Remaining resources will go to the next assignment until the iteration meets the termination.

\section{Performance Evaluation} \label{sec:evacuation}

Detailed setup and results of our experiments are illustrated in this section, which provide a more complete understanding of the efficacy and efficiency of our proposed schema.

\subsection{Implementation Details}

\noindent \textbf{Data source:} We collect flight and human-factor data for 2017 September from the Airline On-Time Performance Database and 123ATC.com\footnote{https://123atc.com/}.

\noindent \textbf{Evacuation Plans:} We assume that residents in Miami (MIA) would like to be evacuated by flight 6 hours before the arrival of the hurricane $T_h$, and there are 5 supporting and destination airports -- ATL, BNA, CLT, DFW, and JFK. The time step in the L2D system is set as 1 hour. 200 evacuation flights are generated at each time step, and the evacuation plan must be completed before $T_h$ with a fixed number of evacuation-oriented aircraft $N$, which are evenly assigned to the relevant airports. We also suppose that aircraft can transport people immediately without waiting for overhaul, and passenger boarding and landing. 

\begin{figure*}
    \centering
    \begin{subfigure}[b]{0.3\textwidth}
        \centering
        \includegraphics[scale = .32]{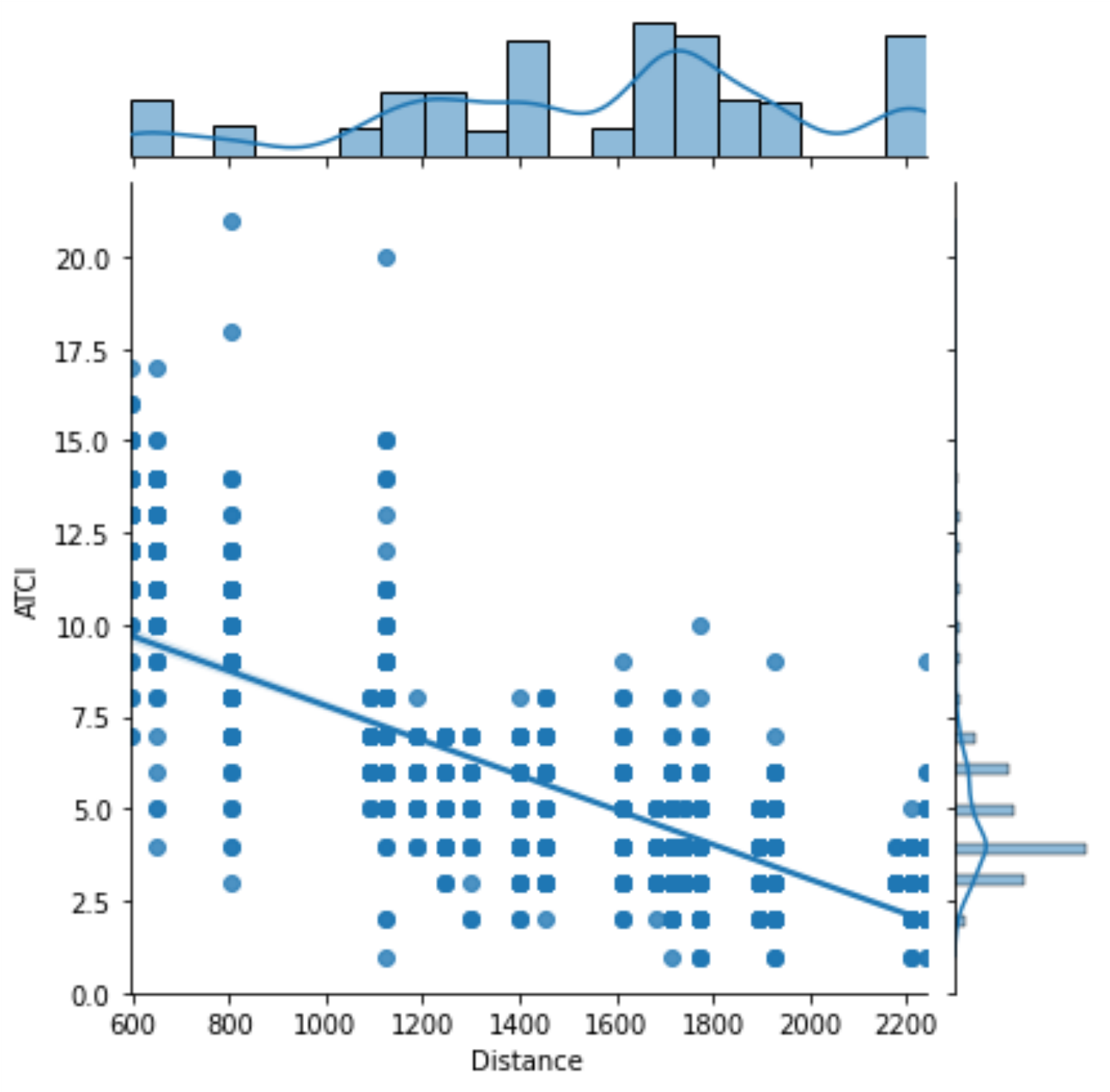}
        \caption{}
    \end{subfigure}
    \begin{subfigure}[b]{0.3\textwidth}
        \centering
        \includegraphics[scale = .32]{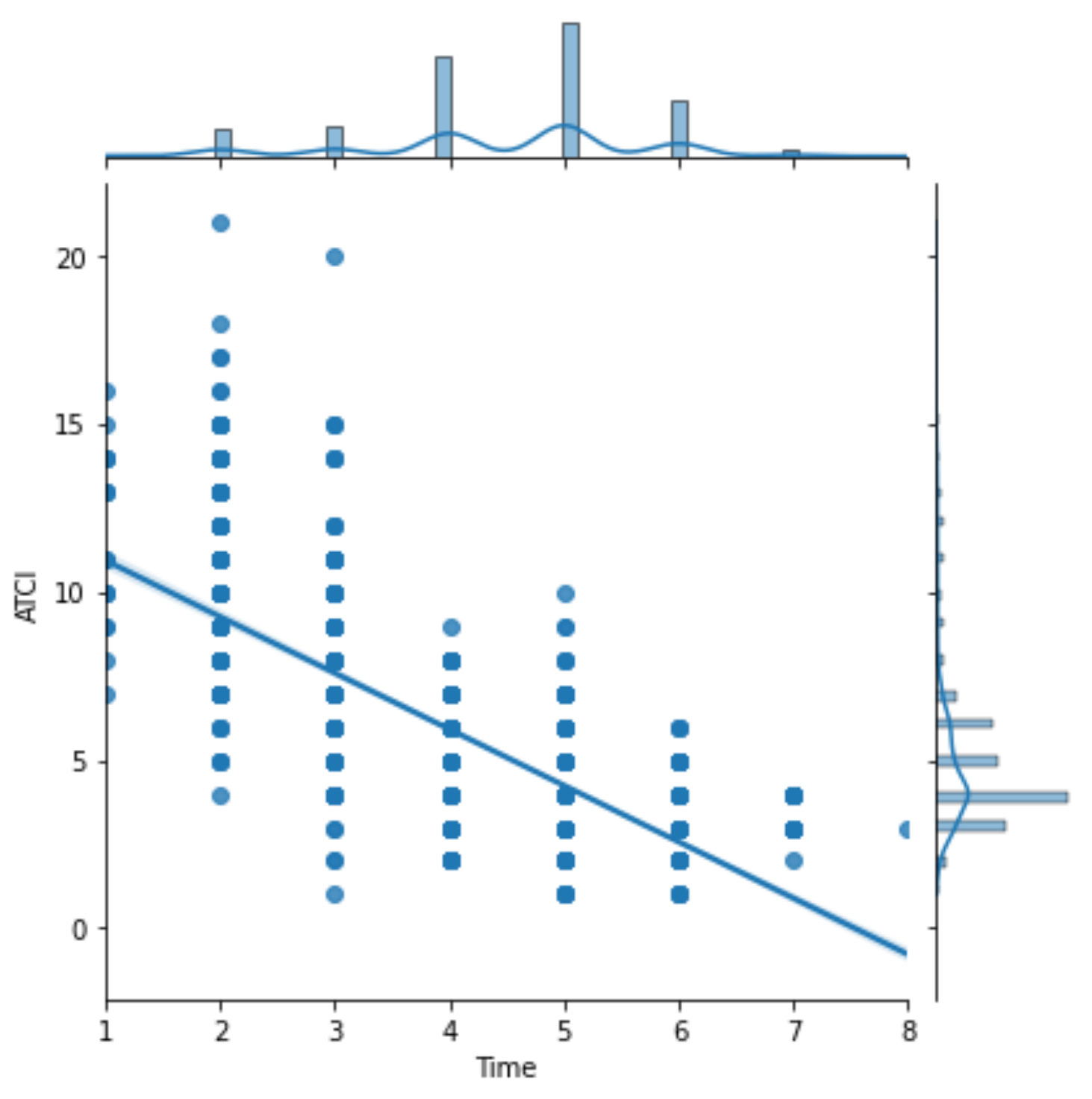}
        \caption{}
    \end{subfigure}
    \begin{subfigure}[b]{0.3\textwidth}
        \centering
        \includegraphics[scale = .32]{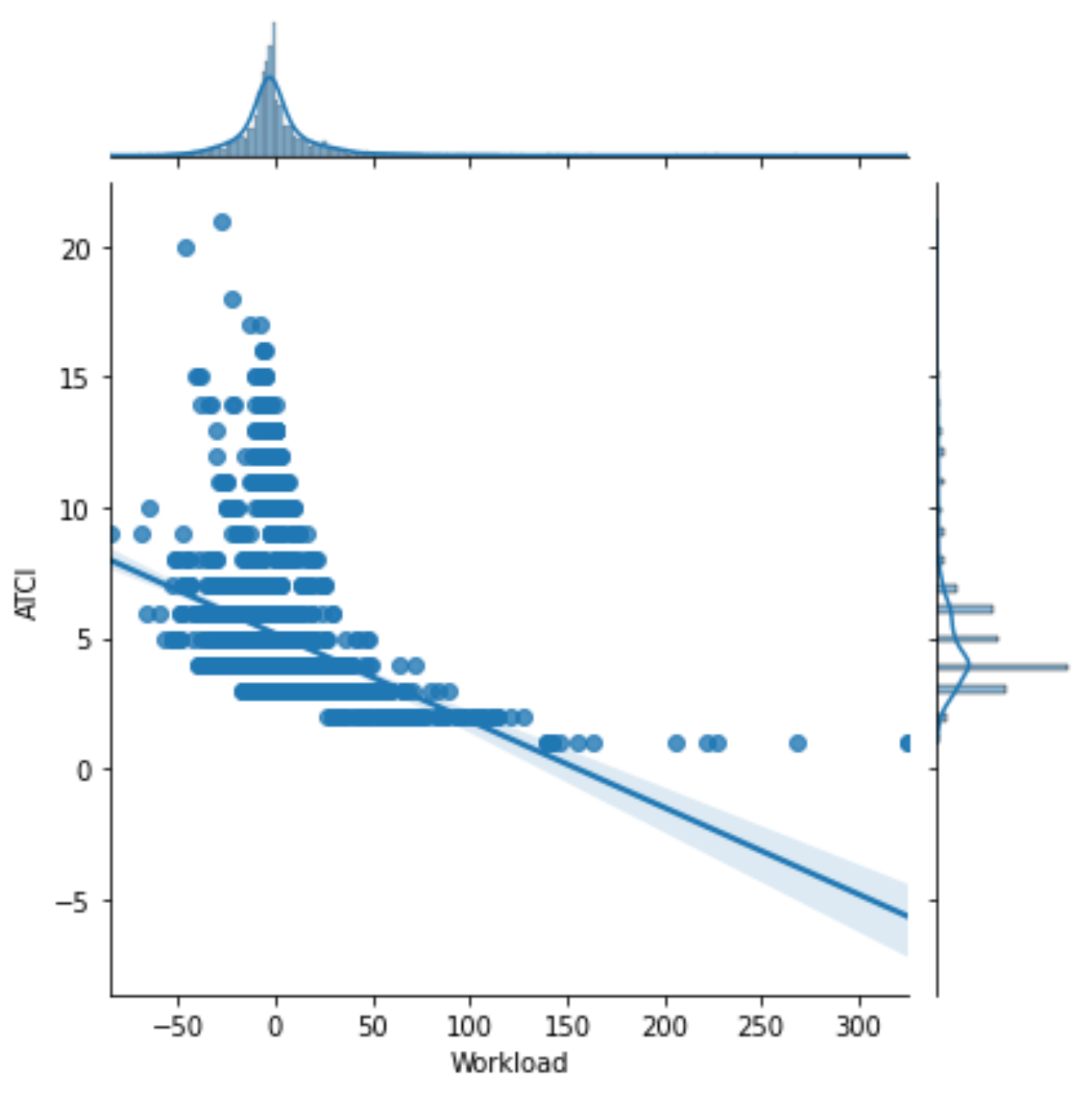}
        \caption{}
    \end{subfigure}
    \caption{The distribution and relationships amongst multiple indexes.}
    \label{fig:correlation}
\end{figure*}

\noindent \textbf{Reward function settings:} Table \ref{tab:hyper} shows some pre-defined hypterparameter candidates and their Spearman rank-order correlation coefficients in terms of ATCI. As described in Algorithm \ref{alg:hyper_selection}, we select $\lambda = 1, \mu = 0.01, \nu = 0.2$ in our reward function design. The correlations of indexes for the optimal hyperparameter tuple are depicted in Figure \ref{fig:correlation}. We can see that $w, et, d$ are all roughly negatively correlated with the ATCI, which tallies with the reward function design intention.

\begin{table}
\caption{Hyperparameters and Correlation Coefficients}
\centering
\begin{tabular}{lccc}
\toprule 
\textbf{Candidate $\langle \lambda^*, \mu^*, \nu^* \rangle$}      & \multicolumn{1}{c}{\textbf{$\rho^t$}} & \multicolumn{1}{c}{\textbf{$\rho^d$}} & \multicolumn{1}{c}{$\rho^w$} \\ \midrule 
\textbf{(1,0.01,0.2)}    & \textbf{-0.77}                         & \textbf{-0.73}                         & \textbf{-0.44}                             \\ 
(1,0.01,0.1)    & -0.89                         & -0.84                         & -0.2                              \\
(1,0.01,0.5)    & -0.45                         & -0.5                          & -0.63                             \\
(1,0.005,0.2)   & -0.66                         & -0.6                          & -0.5                              \\
(1,0.006,0.2)   & -0.65                         & -0.65                         & -0.5                              \\
(1.5,0.006,0.2) & -0.72                         & -0.73                         & -0.42                             \\
(1.5,0.006,0.4) & -0.47                         & -0.5                          & -0.66                             \\ \bottomrule
\end{tabular}

\label{tab:hyper}
\end{table}

\noindent \textbf{Evaluation metrics:} Our pre-defined ATC impact factor is the multiplicative inverse for workload, flight length and elapsed time, thus the larger it is, the lower incremental air traffic complexity is. We also consider two more intuitive evaluation metrics: \textit{"How many evacuation flights can be completed within 6 hours?"} and \textit{"How many evacuation flights each airport has to coordinate with?"} 

\subsection{Hurricane Irma Example}

The eye of Irma is close to Southern Florida in the morning on September 10, 2021. To coordinate evacuation flights in a safer manner, we set 6 AM on that day as the evacuation end time. $N$ is set as 30, 60, 120 respectively. All other experimental settings have been described in the implementation details. 

\begin{figure*}
    \centering
    \includegraphics[scale = .35]{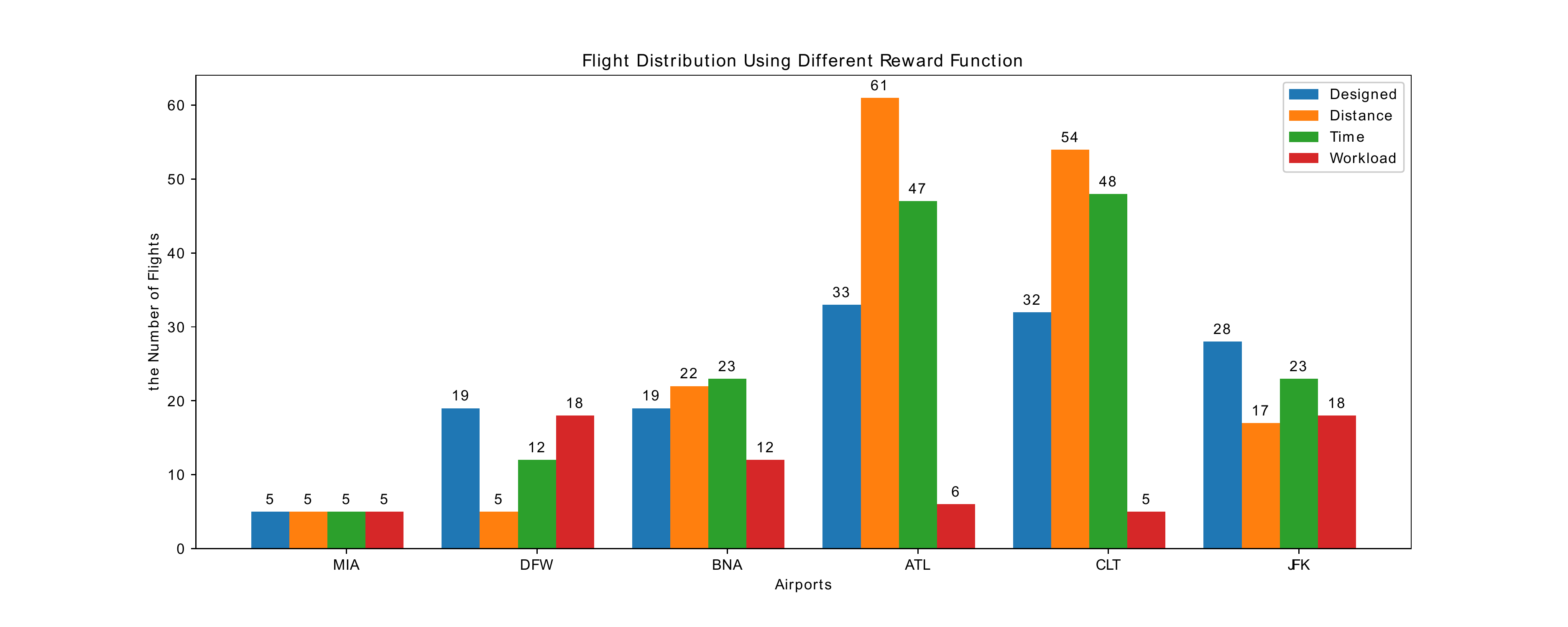}
    \caption{Flight distribution with different reward functions. The transit flights via MIA are not counted while $N = 30$.}
    \label{fig:compar1}
\end{figure*}

\begin{figure*}
    \centering
    \includegraphics[scale = .3]{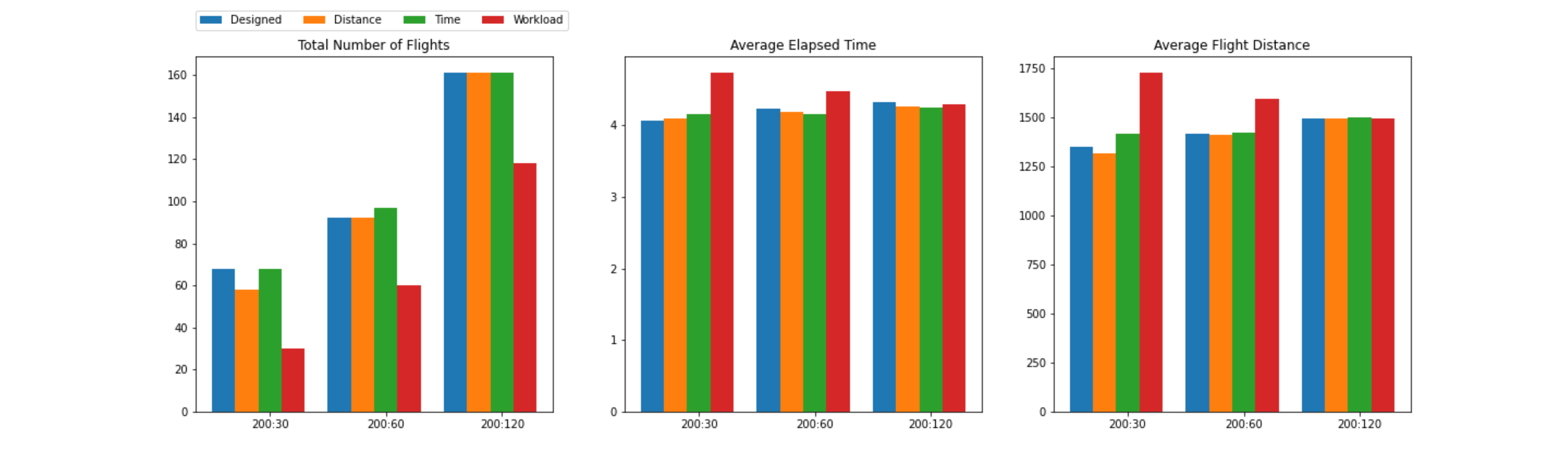}
    \caption{Comparison of distance-based method, time-based method, workload-based method and the ATCI-based method in three metrics on the Hurricane Irma example. X-axis stands for the ratio of potential airway-aircraft at each time step.}
    \label{fig:compare2}
\end{figure*}

Figure. \ref{fig:compar1} shows comparison results of the four methods for the number of completed flights within 6 hours starting at 12 AM on September 10, 2021. Note that the transit flights via MIA are not counted and there are a total of 5 departure flights. It is not astonishing that the time-based and distance-based methods tend to dispatch evacuation flights to ATL and CLT since they are the closest airports to MIA compared to the other 4 airports, while the elapsed time is usually positively correlated with flight length. On the contrary, very few evacuation flights will take off or land from ATL and CLT using the workload-based method. Such a result reconciles with the real conditions since they are two of the world's busiest airports as well as the nearest airports to MIA, evacuation flights could increase airspace complexity in a short time. Besides, airports in the experimental settings are almost large hub/core airports in the US. Therefore, it is easy to explain that why the workload-based method produces far fewer airway-aircraft pairs than others, the same result is also shown in Figure. \ref{fig:compare2}. It is worth pointing out that a more balanced flight allocation plan is obtained with our designed reward function. The extra workload caused by evacuation is nearly uniformly shared by all airports.

The comparison results for different airway-aircraft ratios are illustrated in Figure. \ref{fig:compare2}. Here the "200" in the airway-aircraft ratio indicates that 200 potential airways are generated at each time step, then they can be paired with available aircraft. As the ratio increases, there is a convergent phenomenon in the evaluation metrics depending on average elapsed time and flight length. In other words, all strategies tend to achieve similar performance because of sufficient aircraft supply. Otherwise, when the supply is not enough such as $N = 30$ or $N = 60$, the workload-based method results in more flight time and flight distance for each evacuation flight compared to the other three methods. 

Assume that an evacuation aircraft can transport 400 people at a time, approximately 28,000 people can evacuate towards other cities within 6 hours based on L2D even if only 30 aircraft serves for evacuation. Therefore, Figure. \ref{fig:compar1} and Figure. \ref{fig:compare2} jointly state that the ATCI-based method can produce good and unbiased evacuation flight plans.

\section{Conclusion} \label{sec:conclusion}
In this paper, we propose a new flight dispatch system L2D under emergency that aims to optimize the air route network's long-term efficiency, as well as satisfying instant evacuation demands. The flight dispatch is modeled as an MDP, where the value of the aircraft-airway pair is obtained by the flight's utility and the future expected ATCI value learned from synthetic data based on historical records. Matching between multiple aircraft and potential evacuation airways is then solved by the Hungarian algorithm. The real-world case study based on Hurricane Irma reveals the effectiveness and efficiency of the proposed L2D and ATCI-based reward function compared to the baseline univariate methods. Although it is an early work on flight dispatch under emergency, the results show its potential to be deployed in the real-world disaster management system in the future.

Our proposed L2D is still in the early stage. For future work, we are interested in developing a large-scale and real-time flight dispatch schema based on passenger demand and weather fluctuation, which could be applied for nationwide evacuation planning. Meanwhile, it is beneficial to investigate the approach which can better capture the dynamics of the air route network during disaster evacuation.















\bibliographystyle{./bibliography/IEEEtran}
\bibliography{./bibliography/IEEEabrv,./bibliography/IEEEexample}


\end{document}